 \documentclass[smallabstract,smallcaptions]{dccpaper}

\usepackage{epsfig}
\usepackage{citesort}
\usepackage{amsmath}
\usepackage{amssymb}
\usepackage{color}
\usepackage{url}
\usepackage{subfigure}
\usepackage{amsfonts}
\usepackage{algorithmic}
\usepackage{algorithm}
\usepackage{array}
\usepackage{textcomp}
\usepackage{stfloats}
\usepackage{url}
\usepackage{verbatim}
\usepackage{graphicx}
\usepackage{cite}
\usepackage{multirow}
\newcommand{\etal}{\textit{et al.}}

\newlength{\figurewidth}
\newlength{\smallfigurewidth}

\begin{document}

\title
{\large
\textbf{Accelerating Block-level Rate Control for Learned Image Compression}
}

\author{%
Muchen Dong, Ming Lu, and Zhan Ma\\[0.5em]
{\small\begin{minipage}{\linewidth}\begin{center}
\begin{tabular}{c}
Nanjing University
\end{tabular}
\end{center}\end{minipage}}
}

\maketitle
\thispagestyle{empty}

\begin{abstract}
Despite the unprecedented compression efficiency achieved by deep learned image compression (LIC), existing methods usually approximate the desired bitrate by adjusting a single quality factor for a given input image, which may compromise the rate control results. Considering the Rate-Distortion ($R-D$) characteristics of different spatial content, this work introduces the block-level rate control based on a novel $D-\lambda$ model specific for LIC. Furthermore, we try to exploit the inter-block correlations and propose a block-wise $R-D$ prediction algorithm which greatly speeds up block-level rate control while still guaranteeing high accuracy. Experimental results show that the proposed rate control achieves up to $100\times$ speed-up with more than 98\% accuracy. Our approach provides an optimal bit allocation for each block and therefore improves the overall compression performance, which offers great potential for block-level LIC.  

\end{abstract}

\section{Introduction}

Recent years, learned image compression (LIC)~\cite{balle2016end,balle2018variational,minnen2018joint,lu2022high} has witnessed incredible development with superior compression performance, even surpassing the traditional VVC intra profile~\cite{vvc_intra}. With the data-driven end-to-end joint optimization, such learning-based codecs have achieved the optimal rate-distortion ($R-D$) result at each bitrate point, which is controlled by an additional Lagrange coefficient $\lambda$. Typically, existing methods attempt to approach different bitrates by training multiple fixed rate models or the variable-rate counterpart~\cite{chen2020_variable}. Nevertheless, such a way to decide an appropriate $\lambda$ value in image level by iterative coding processes significantly hinders the practical application, which not only leads to time wasting, but also degrades the bit allocation.
In principle, large content variations are presented on different image blocks. Texturally complex and smooth regions tend to exhibit different $R-D$ characteristics. As a result, it is crucial to model the block-level $R-D$ behaviors and conduct optimal rate control for each block accordingly.


Rate control is designed to achieve the target bitrate while minimizing compression distortion. Currently, there are few studies related to rate control for LIC at block level. Cai \etal~\cite{cai2018efficient} introduced a prediction network on top of the image-level compression model to predict the specific bitrate allocated to each spatial block. This method relied on the learning-based prediction from the whole image without considering the inherent $R-D$ relationship, which was impractical for block-based coding. Wang \etal~\cite{wang2022block} proposed an enumeration-based approach to determine the concrete bitrate of each block as well as to achieve the overall target bitrate with minimized distortion. Although this methodology allowed for precise rate control results, it required repeated coding of each block for $R-D$ modeling, which is unacceptably time-consuming. 


To address the above issues, we propose an efficient and fast block-level rate control for LIC. Our algorithm is implemented in two steps: 1) Firstly, we define a novel $D-\lambda$ function specific for LIC in accordance with the normalized Lagrange factor $\lambda$. This leads our model to achieve more accurate fitting results and eventually benefit the rate control performance. 2) Secondly, to avoid the inefficiency of $R-D$ modeling block-by-block, we try to exploit the texture correlations across spatial blocks. Owing to our modified relations between $\lambda$ and $D$, we establish linear functions between image gradients and model parameters for any spatial content. By sampling on a small number of blocks, the $R-D$ curves of remaining blocks can be directly obtained without any coding processes, which greatly reduces the time consumption of rate control while also ensures high accuracy.

We conduct evaluations on the widely used Kodak and Tecnick datasets. Experimental results demonstrate that our proposed method greatly speed up the rate control process by up to $100\times$ compared to previous methods and still with a high accuracy of 98\%. By means of our rate control algorithm, we can additionally achieve optimal bit allocation and thus improve the overall compression performance of block-level LIC. 


\section{Related Works}

\subsection{Learned Image Compression}

Learned image compression (LIC) methods~\cite{balle2016end,minnen2018joint,cheng2020learned,lu2022high} have made significant progress in the last few years. Most of them are dedicated to deploying more and more sophisticated transforms or context models to improve the overall coding efficiency, which leads to increasing computational consumption and decoding complexity for the practical application. Consequently, the practical image compression techniques, such as model simplification and parallel decoding acceleration, have recently become a major research area of interest in the academic community. Block-level LIC~\cite{wu2021learned} is one of the solutions to alleviate the memory usage by uniformly slicing the input image into blocks with small size. Though promising improvements achieved by block-based techniques (e.g., inter-block prediction), few works have considered block-level rate control, resulting in suboptimal bit adaptation and inferior compression efficiency.



\subsection{Rate Control} \label{sec:related_rate_control} 

With the exception of individual efforts~\cite{cai2018efficient} to directly predict bit allocation using neural networks, the concept of rate control is basically built upon the $R-D$ modeling. Studies on the relationship between $R$ and $D$ have been uninterrupted, of which the most representative are the exponential model~\cite{sullivan1998rate} and the hyperbolic model~\cite{mallat1998analysis,dai2004rate}. At the same time, the rate control methods of traditional codecs can also be generally categorized into $QP$-domain, $\rho$-domain, and $\lambda$-domain, respectively. Regarding learning-based codecs, the rate control is usually done in $\lambda$-domain as the supervised training by the $R-D$ loss function. Jia \etal~\cite{jia2022rate} proposed the first image-level rate control for LIC by means of $R-D$ modeling. Apart from that, most existing works approached the desired bitrate by iteratively adjusting the $\lambda$ for a given image on the basis of pre-trained models, and the block level as well. Wang \etal~\cite{wang2022block} firstly considered the block-level rate control for LIC, which achieved the optimal bit allocation at the target bitrate by traversing the bit steps for each block. Although high accuracy has been achieved by this method, it was very time-consuming to obtain the $R-D$ relationship and unfavorable for practical application.

\section{Proposed Method}

\subsection{Problem Statement}

Generally speaking, the problem of rate control in block level can be divided into two steps: 1) modeling the accurate rate-distortion relationship for each specific block; 2) deriving approximate $\lambda$ per block to minimize the overall distortion as below:


\begin{align}
\{Para\}_{opt} =\mathop{\arg\min}\limits_{\{Para\}}\sum_{i=0} ^{N} D_i,
\label{eq:block-level-BA-problem}
\quad \text{s.t.} \sum_{i=1}^{N} R_{i} \leq R_T.
\end{align}
$N$ is the total number of blocks to be encoded. the distortion $D_i$ is induced by the lossy coding of the $i$-th block, often measured by the mean square error (MSE), while $R_i$ is the corresponding bitrate. $R_T$ represents the overall target bitrate. Having the optimized set of $\{ \lambda_1, \lambda_2, ..., \lambda_N \}_{opt}$ to determine the coding parameters $\{Para\}_{opt}$ for each block, we can achieve the minimized overall distortion at a given bitrate.


In order to achieve accurate $R-D$ modeling, although the traversal coding in \cite{wang2022block} provides a straightforward solution, it is difficult to apply in practice, especially for such block-level task. In this work, we turn to utilize the underlying statistical $R-D$ characteristics to accelerate the modeling process, by which we only need several coding operations to calculate the $R-D$ model coefficients to obtain the entire functional curves.


As we disscussed in Sec.~\ref{sec:related_rate_control}, there have been a lot of successful researches on $R-D$ modeling. The exponential fucntion~\cite{sullivan1998rate} is one of the classic models which is constructed as:
\begin{align}
    D = C\textit e ^{-KR},
    \label{eq:exp_RD_model}
\end{align}
where $C (C \geq 0)$ and $K (K \geq 0)$ are model parameters related to the content of images and can be calculated by coding at least twice. Noted that these parameters determine the overall fitting accuracy and can be improved by evaluating at the close bitrate or with similar texture content. From the perspective of $R-D$ loss, $\lambda$ can be seen as the slope of the $R-D$ curve which can be derived from:
\begin{align}
\lambda = -\frac{\partial D}{\partial R} = CKe^{-KR}.
\end{align}
Therefore we can obtain the $R-\lambda$ function by:
\begin{align}
\label{eq:exp_lambda_r}
R = -\frac{1}{K}ln \lambda + \frac{1}{K}ln(CK) = a ln \lambda + b,
\end{align}
where $a = -\frac{1}{K}$ and $b = \frac{1}{K}ln(CK)$.


Based on Eq. \ref{eq:exp_RD_model} and Eq. \ref{eq:exp_lambda_r}, the relationship between $D$ and $\lambda$ can be derived as:
\begin{align}
\label{eq:exp_lambda_d}
&D = Ce^{-KR} = Ce^{-K(a ln \lambda + b)} = Ce^{-Kb}\lambda^{-K a }, \nonumber \\
&\implies D = c \lambda ^{d}.
\end{align}

\subsection{Reformulation of $D-\lambda$ Model for LIC}
\label{subsection:RDmodel}

\begin{figure*}
	\centering
        \subfigure[kodim01]{
            \begin{minipage}[b]{0.45\textwidth}
            \includegraphics[width=1\textwidth]{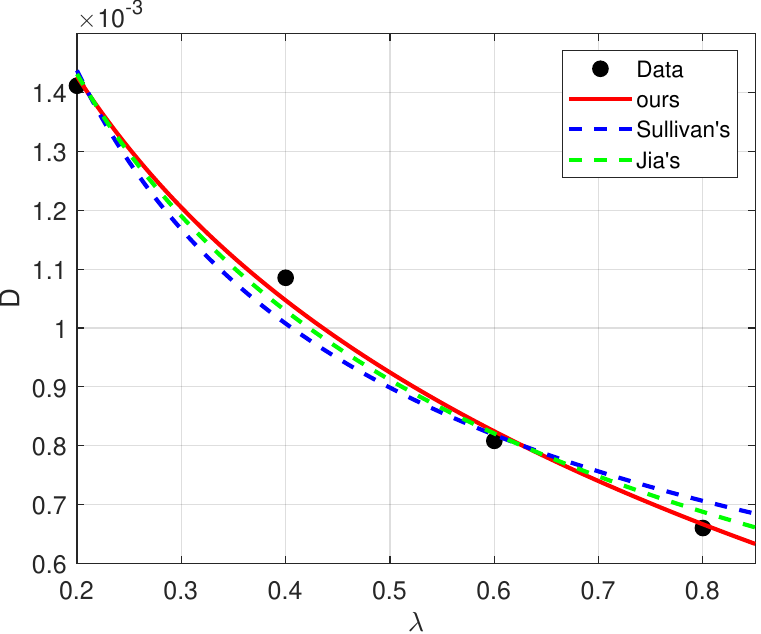}
            \end{minipage}
        \label{fig:fit_compare_4}
        }
        \subfigure[kodim13]{
		\begin{minipage}[b]{0.45\textwidth}
		\includegraphics[width=1\textwidth]{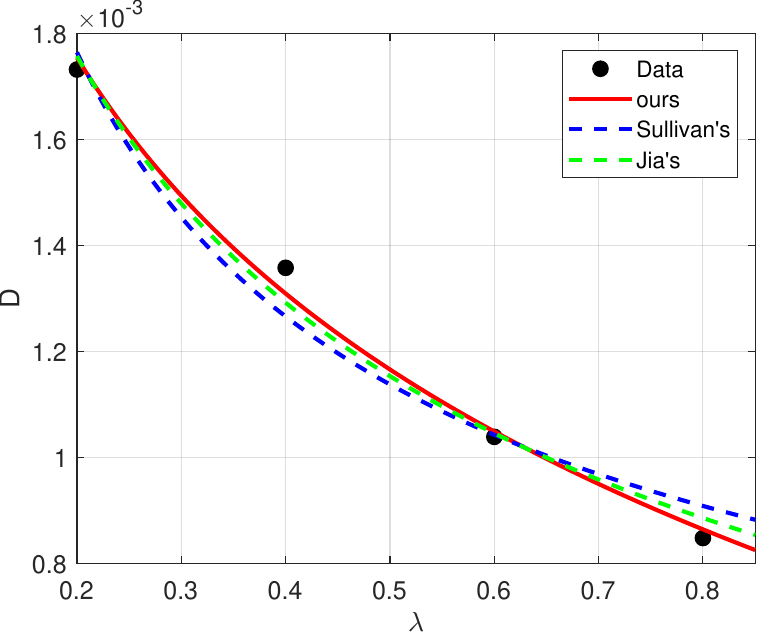} 
		\end{minipage}
		\label{fig:fit_compare_5}
	}
	\caption{Fitting comparison results of $D-\lambda$ model of Sullivan \etal \cite{sullivan1998rate} (\textbf{\textcolor{blue}{blue}}), Jia \etal \cite{jia2022rate} (\textbf{\textcolor{green}{green}}) and ours (\textbf{\textcolor{red}{red}}). Our model proves the best accuracy.}
	\label{fig:fit_compare}
\end{figure*}

Recalling the $D-\lambda$ model in Eq.~\ref{eq:exp_lambda_d}, once the hyperparameters are fixed, the magnitude of D depends completely on the variations of $\lambda$, which in traditional codecs are usually positive values that can be as high as hundreds. However, in our implementation of variable-rate model as in \cite{wang2022block}, the $\lambda$ values are usually normalized to between 0 and 1 from the real $\lambda_{rd}$ used for $R-D$ optimization\footnote{$\lambda=1$ usually corresponds to the largest value of $\lambda_{rd}$.}. This prompts us to further adjust the $D-\lambda$ model in Eq.~\ref{eq:exp_lambda_d} to better accommodate changes of $\lambda$. Considering the loss function $R + \lambda_{rd} D$ for supervision of model training, as $\lambda$ tends to 0, the constraint on the $D$ term will then be loosen, which means minimizing the $R$ term will be the main concern during the optimization, thus the $D$ will tend to a positive infinity value. Conversely, when $\lambda$ approaches 1, the training process will make $D$ decreasing, ideally converging to 0. The above on the relationship between $D$ and $\lambda$ is functionally analogous to the negative logarithmic function. As such, we propose a new function to fit the $D-\lambda$ relationship as:
\begin{align}
\label{eq:D_lambda_model}
    D = -m log_t \lambda + n = -m \frac{ln \lambda}{ln t} + n = a' ln \lambda + b',
\end{align}
where we use $e$ as the base for ease of calculation. Since our approach is oriented to the block level, we choose two blocks of size $256\times256$ from the Kodak dataset as samples for fitting evaluation. As shown in Fig.~\ref{fig:fit_compare}, our $D-\lambda$ model demonstrates more accurate than the original model (see Eq.~\ref{eq:exp_lambda_d}) proposed by Sullivan \etal~\cite{sullivan1998rate} and another by Jia \etal \cite{jia2022rate}.

So far, we have obtained relational equations for both $R-\lambda$ and $D-\lambda$ in the same logarithmic formulation, which will help us to quickly obtain $R-\lambda$ and $D-\lambda$ relationships for any spatial contents if the functional coefficients of different blocks are correlated in such form.


\subsection{Accelerating Rate Control with Inter-Block Correlation}
\label{subsection:Boosting rate control}

Typically, given an input block, we only need to deploy the encoding and decoding twice using the pretrained model to get all the coefficients for $R-\lambda$ and $D-\lambda$. However, it is practically unfeasible to conduct the coding processes for each block to get complete $R-D$ curves. In principle, for the same image, there must be a strong texture correlation between different blocks in terms of spatial content, which implies that there is also a strong inter-block correlation regarding $R-D$ relationship. In this work, we utilize the average gradient $\nabla_{pp}$ as an indicator of the texture complexity for a given image by:
\begin{align}
\label{eq:gradient_of_image}
\nabla_{pp} = \frac{1}{HW} (\sum_{i=0}^{W-1} \sum_{j=0}^{H-1} (I_{i,j}-I_{i+1,j})^2 + (I_{i,j}-I_{i,j+1})^2)^{\frac{1}{2}},
\end{align}
where $W$ and $H$ denote the image width and height respectively. $I_{i,j}$ is the pixel value at position $(i,j)$. We deploy observational experiments on the Kodak dataset and Tecnick dataset to model the relationships between the image gradient and the $R-D$ characteristics. Specifically, each input image is divided into non-overlapped $256\times256$ blocks to pass into the coding processes, having the $R-\lambda$ and $D-\lambda$ curves. Then we calculate the average gradient of each block via Eq.~\ref{eq:gradient_of_image} to construct the relations with the model coefficients.

\begin{figure*}
	\centering
	\subfigure[kodim01]{
		\begin{minipage}[b]{\textwidth}
			\includegraphics[width=0.23\textwidth]{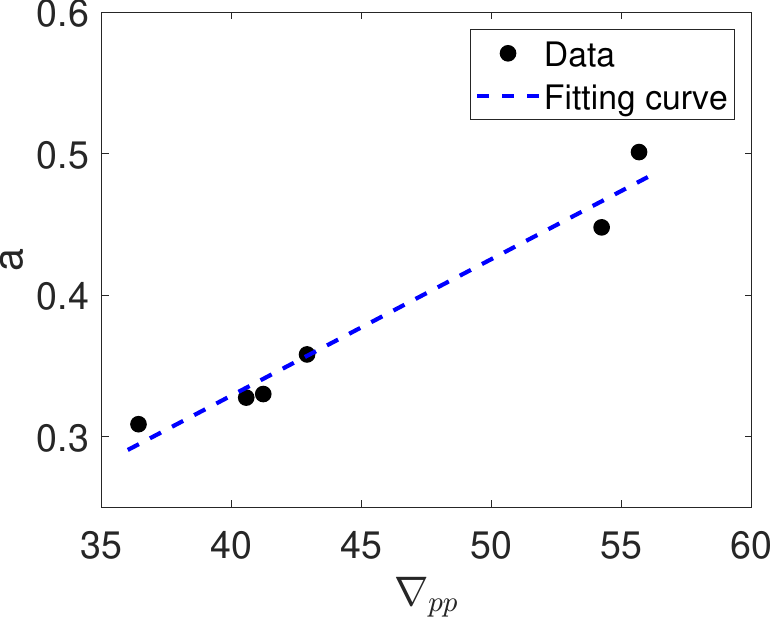}
   			\includegraphics[width=0.24\textwidth]{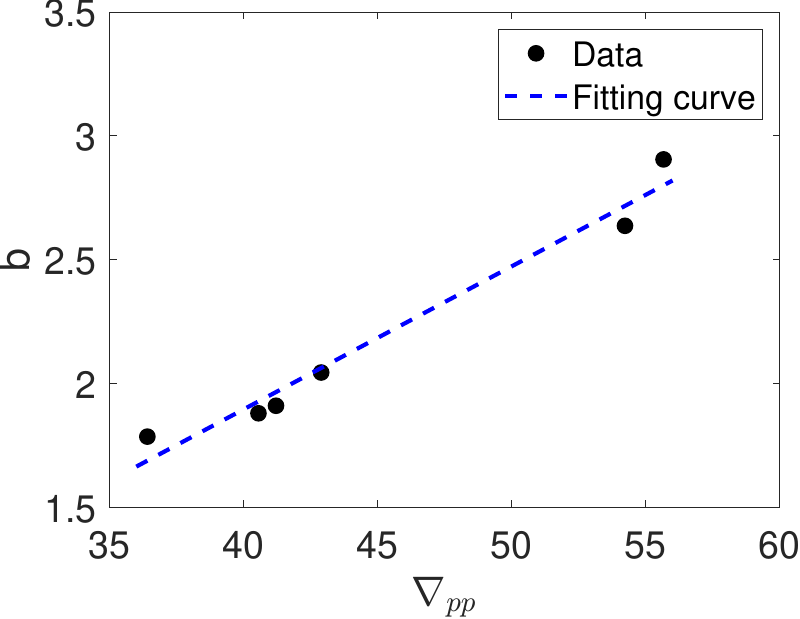}
      		\includegraphics[width=0.24\textwidth]{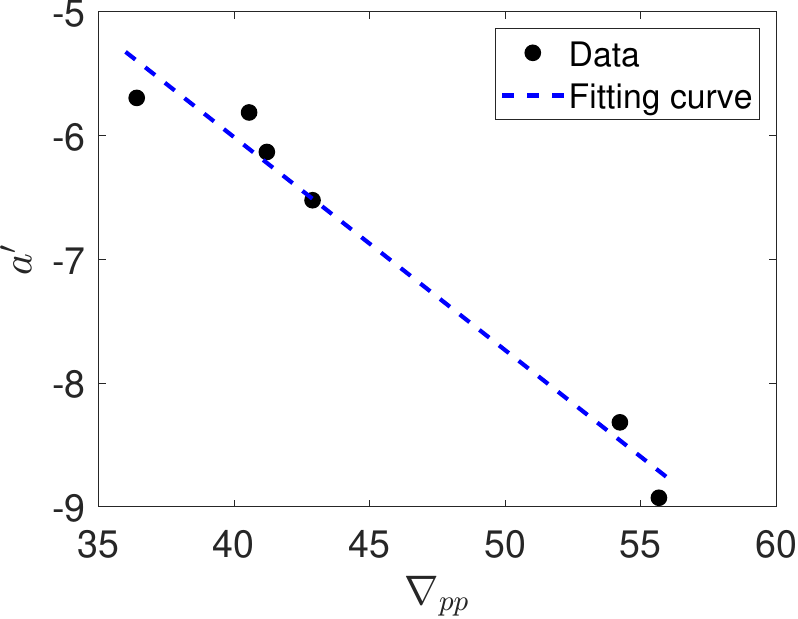}
         	\includegraphics[width=0.24\textwidth]{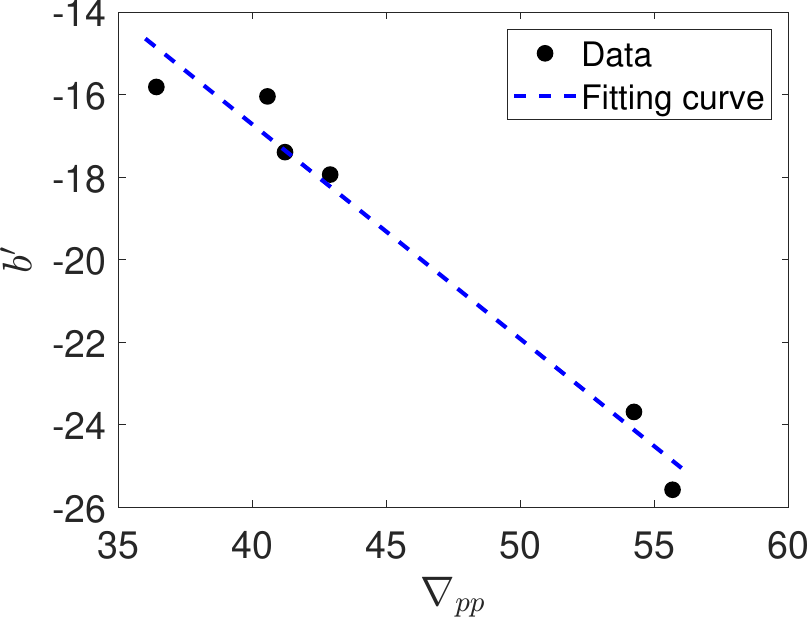}
   
		\end{minipage}
		\label{fig:gradient_fit1}
	}
         \\
        \subfigure[Tecnick50]{
            \begin{minipage}[b]{\textwidth}
                \includegraphics[width=0.24\textwidth]{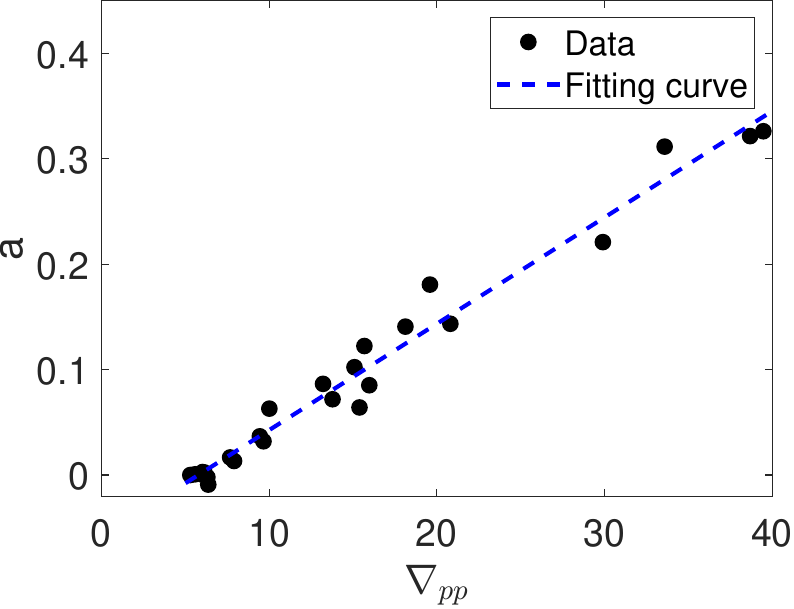}
   			\includegraphics[width=0.23\textwidth]{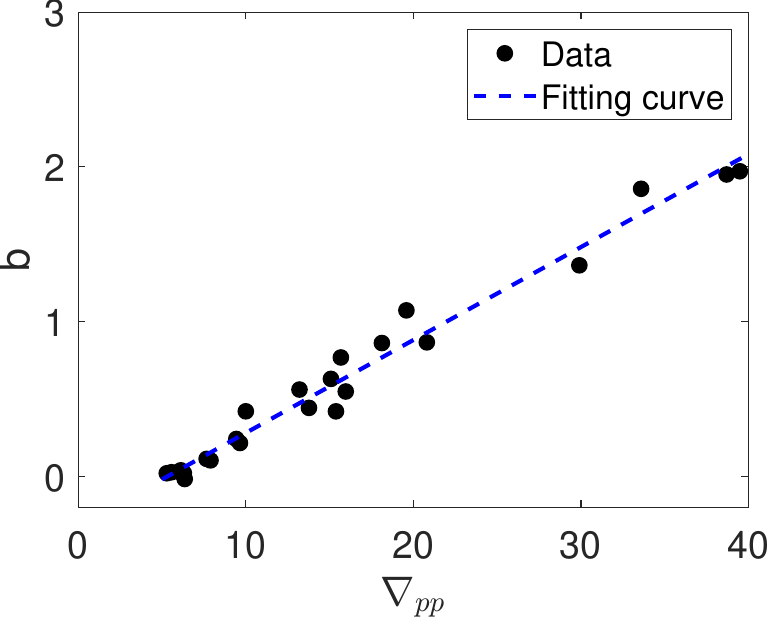}
      		\includegraphics[width=0.24\textwidth]{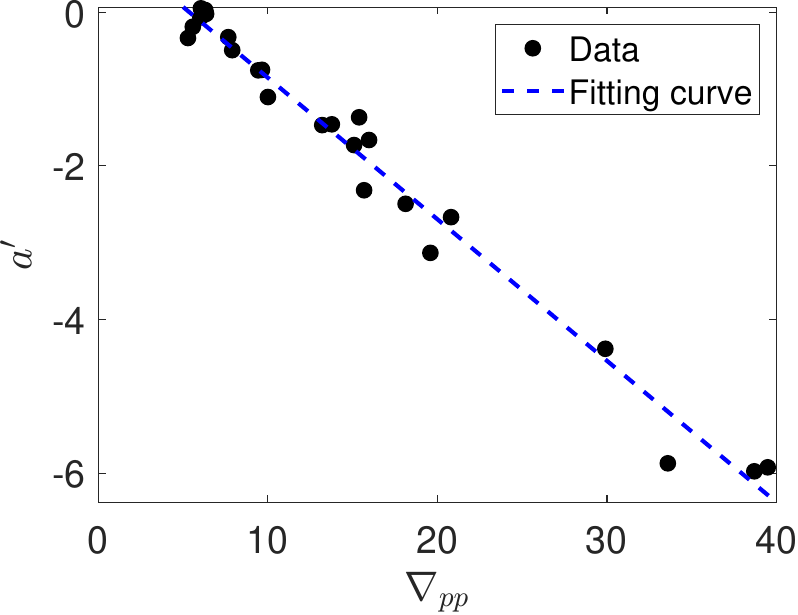}
         	\includegraphics[width=0.24\textwidth]{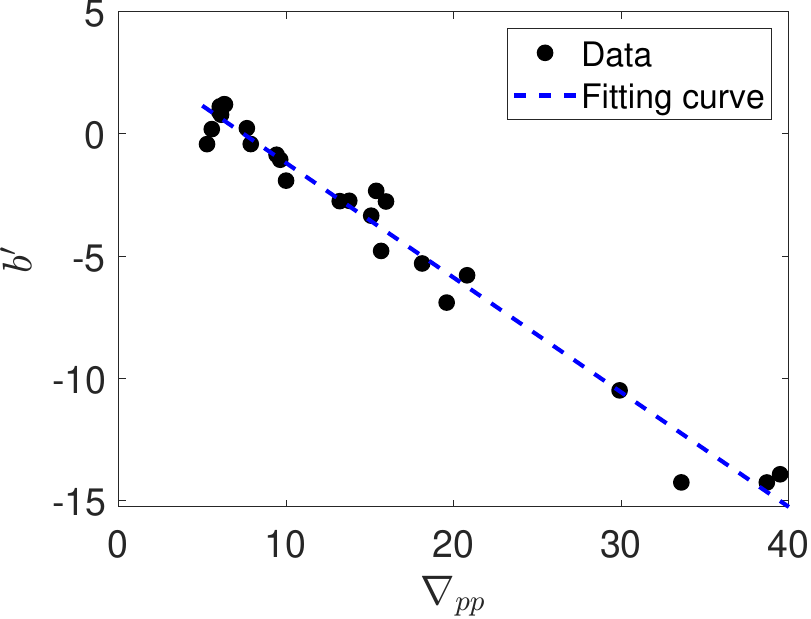}
            \end{minipage}
        \label{fig:gradient_fit4}
        }
	\caption{Fitting results of relationships between $\nabla_{pp}$ and coefficients in our $R-\lambda$ and $D-\lambda$ models.}
	\label{fig:gradient_fit}
\end{figure*}

Figure \ref{fig:gradient_fit} depicts the sampled correlations between the average gradients $\nabla_{pp}$ and the coefficients $a$, $b$ in $R-\lambda$ model as well as $a'$ and $b'$ in $D-\lambda$ model, in which each point corresponds to a specific block. We find a clear linear relationship of blocks with different contents between the average gradient and each coefficient, which means that by calculating the model coefficients for a small number of blocks, the coefficients for the remaining blocks can be determined quickly via corresponding image gradients. This provides a noticeable acceleration for block-wise $R-D$ modeling instead of duplicated encoding and decoding processing.





In specific implementation, we divide the input image without overlapping and sample a number of blocks to perform encoding and decoding processes. By means of twice encoding and decoding for each sampled block, we can calculate the $R-\lambda$ and $D-\lambda$ relationships using Eq. \ref{eq:exp_lambda_r} and Eq. \ref{eq:D_lambda_model}. Furthermore, having the average gradient $\nabla_{pp}$ of each block, the linear relations for all blocks can be determined by the sampled ones. As a result, the $R-\lambda$ and $D-\lambda$ models for unsampled blocks can be predicted without any coding processes.

Once the block-adaptive $R-\lambda$ and $D-\lambda$ models are obtained, we try to optimize the bitrate assigned to each block to minimize the overall distortion for rate control. We first assume the $\lambda$ values of all blocks are initialized to $\lambda_{init}$ which makes the overall bitrate higher than the target, and the adjustment step size is $\lambda_{step}$. We use $R_i$, $D_i$ and $C_i$ to represent the bitrate, distortion, and distortion cost for adjustment of the $i$-th block. The process of block-level rate control is as follows:


Firstly, we calculate the bitrate for each block using $R-\lambda$ model with $\lambda_{i}$. Our goal is to adjust the overall bitrate as close as possible to the target bitrate.


Secondly, we calculate the distortion cost required to adjust the $\lambda_i$ of each block to $\lambda_i - \lambda_{step}$ as below:
\begin{align}
\label{eq:metric_ours}
&C_i =\frac{a'_i}{a_i}(R_i(\lambda_i)-R_i(\lambda_i-\lambda_{step}))
=a'_iln\frac{\lambda_i}{\lambda_i - \lambda_{step}},
\end{align}
$a'_i$ and $a_i$ are the $D-\lambda$ and $R-\lambda$ coefficients of the $i$-th block, respectively.

Thirdly, after obtaining the cost for each block, we select the block with the smallest cost to update $\lambda_i$ to $\lambda_{i}-\lambda_{step}$ with refined bitrate. 

Finally, we update the overall bitrate consumption based on the Eq. \ref{eq:exp_lambda_r}. If the total bitrate is less than $R_T$ for the first time, we output the list that records the $\lambda_i$ of each block. Otherwise, we jump to the second step and repeat the following steps until the total bitrate is less than or equal to the target bitrate $R_T$.




\begin{algorithm}
	\renewcommand{\algorithmicrequire}{\textbf{Input:}}
	\renewcommand{\algorithmicensure}{\textbf{Output:}}
	\caption{Our Rate Control Method} 
        \label{al:our_rate_control}
	\begin{algorithmic}
            \REQUIRE $\lambda_{init}$, $\lambda_{step}$, $R_T$
            \STATE Initialise the $\lambda$ values of all blocks to $\lambda_{init}$.
            \STATE Divide the input image into non-overlapping blocks and sample some blocks.
            \STATE By means of twice encoding and decoding for each sampled block, calculate $R-\lambda$ and $D-\lambda$ relationships for these sampled block using Eq. \ref{eq:exp_lambda_r} and Eq. \ref{eq:D_lambda_model}.
            \STATE Calculate the average gradient $\nabla_{pp}$ of each block using Eq. ~\ref{eq:gradient_of_image}.
            \STATE Determine the linear relations between coefficients and the average gradient $\nabla_{pp}$.
            \STATE Predict the $R-\lambda$ and $D-\lambda$ relationships for unsampled blocks based on the linear relations.
		  \STATE Calculate the rate of each block $R_i$ using Eq. \ref{eq:exp_lambda_r}.
            \WHILE{$\sum_{i=0}^{N} R_i > R_T$}
            \STATE Calculate the distortion cost $C_i$ required to adjust the $\lambda_i$ of each block to $\lambda_i-\lambda_{step}$ using Eq. \ref{eq:metric_ours}.
            \STATE Select the block with the smallest cost to update $\lambda_i$ to $\lambda_i$ - $\lambda_{step}$.
            \STATE Update the bitrate consumption of the selected block using Eq. \ref{eq:exp_lambda_r}.
            \ENDWHILE
            \ENSURE The list that records the $\lambda_i$ of each block.
	\end{algorithmic} 
\end{algorithm}









For a more comprehensive presentation, our rate control algorithm is summarized in Algorithm \ref{al:our_rate_control}.


\section{Experiemnts}

\subsection{Experimental Setup}

We perform the proposed rate control algorithm and other comparison methods using the variable-rate version of TinyLIC~\cite{lu2022high}, which is lightweight with pretty good compression efficiency. To evaluate the performance of rate control, we first set an initialized $\lambda$ value for encoding, resulting in the overall bitrate $R$ measured by bit per pixel (bpp). Then, we set the target bitrate to be 0.95 times of $R$, denoted as $R_T$. For different bitrates range cover, we further set the $\lambda$ values to 0.3, 0.6, and 0.9 for low bitrate, medium bitrate and high bitrate, respectively. Then we conduct corresponding experiments on the Kodak dataset and Tecnick dataset.

\subsection{Performance}

A variety of algorithms are chosen for comparison. We reproduce the models proposed by Sullivan \etal~\cite{sullivan1998rate}, Mallat \etal~\cite{mallat1998analysis}, Jia \etal~\cite{jia2022rate} and Wang \etal~\cite{wang2022block} for rate control at block level. It should be noted that the performance of our model is conditioned on the sampling ratio of blocks to be encoded, and we provide two versions, i.e., the model with the most accuracy (``Ours(Best)'') and the model with the fastest runtime (``Ours(Fast)''). 

We detail the accuracy of rate control and running time required for each method. The results are shown in Table~\ref{table:rate control results}. As can be seen, our model oriented to accuracy of rate control demonstrates the best on all the datasets and bitrate ranges with the error less than 1\%. Even compared with the exhaustive encoding method by Wang \etal~\cite{wang2022block}, our method still leads in accuracy while largely reduces the time consumption for rate control. In addition, our model oriented to speed of rate control dramatically accelerates the rate control process, i.e., 10$\times$ faster on Kodak and 100$\times$ faster on Tecnick when compared to Wang \etal's model~\cite{wang2022block}, and still maintains an accuracy of more than 98\%.

\begin{table}[htpb]
\setlength{\tabcolsep}{3pt}
\caption{Rate control comparisons. Red/Blue Text: Best/Second best Performance.}
\label{table:rate control results}
\centering
\begin{tabular}{cccccccc}
\hline
 \multirow{2}{*}{Dataset}& \multirow{2}{*}{Method}& \multicolumn{2}{c}{low bpp} & \multicolumn{2}{c}{middle bpp} & \multicolumn{2}{c}{high bpp} \\
 &  & $\Delta$R(\%)       & Time(s)  & $\Delta$R(\%)       & Time(s)  & $\Delta$R(\%)        & Time(s) \\ \hline
\multirow{6}{*}{Kodak}  & Sullivan \etal\cite{sullivan1998rate}        & 3.14  &2.845  & 1.85  & 2.804 & 2.48& 2.968  \\
                         & Mallat \etal\cite{mallat1998analysis}      & 2.98  &2.817  & 1.34  & 2.870 & {1.12}& 2.941  \\
                         & Wang \etal\cite{wang2022block}           & \textcolor{blue}{0.62}  &7.981  & \textcolor{blue}{1.00} & 7.226& \textcolor{blue}{0.86}& 8.195   \\
                        & Jia \etal\cite{jia2022rate} &\textcolor{red}{0.29}   &2.606  &1.40  &2.631  &1.44 &2.592   \\                      
                         & Ours(Best)             & \textcolor{red}{0.29}  &\textcolor{blue}{2.333}  & \textcolor{red}{0.74}  &\textcolor{blue}{2.512} & \textcolor{red}{0.50}& \textcolor{blue}{2.400}  \\
                         & Ours(Fast)              & 1.22  &\textcolor{red}{0.845}  & 1.64  & \textcolor{red}{0.844}  & 1.32&\textcolor{red}{0.832}   \\ \hline
\multirow{6}{*}{Tecnick} & Sullivan \etal\cite{sullivan1998rate}       & 4.12  &10.383  & 1.27 & 10.271 & 1.08&10.58  \\
                         & Mallat \etal\cite{mallat1998analysis}      & 3.88  &{10.351}  & {0.83} & 10.492 & {0.74}&10.961     \\
                         & Wang \etal\cite{wang2022block}           & {0.89}  &150.722  &\textcolor{blue}{0.25}  & 117.156  & \textcolor{red}{0.25} & 177.538    \\
                           & Jia \etal\cite{jia2022rate}&\textcolor{blue}{0.77}&11.900  &0.71  &11.296  &0.90 &11.359   \\   
                           & Ours(Best)              & \textcolor{red}{0.66}  &\textcolor{blue}{9.516}  &\textcolor{red}{0.24} & \textcolor{blue}{9.473} & \textcolor{red}{0.25}&\textcolor{blue}{10.533}    \\
                         & Ours(Fast)            &{1.18}  &\textcolor{red}{1.627}  & {0.88}  & \textcolor{red}{1.786}  &\textcolor{blue}{0.72}  & \textcolor{red}{1.793}   \\ \hline
\end{tabular}
\end{table}

\subsection{Block-level vs Image-level}
\begin{figure}[htpb]
	\centering
        \subfigure[0.521bpp 32.47dB $\Delta R$=0.06\%]{
            \begin{minipage}[b]{0.4\textwidth}
            \label{fig:block_level_rate_bit_allocation}
            \includegraphics[width=1\textwidth]{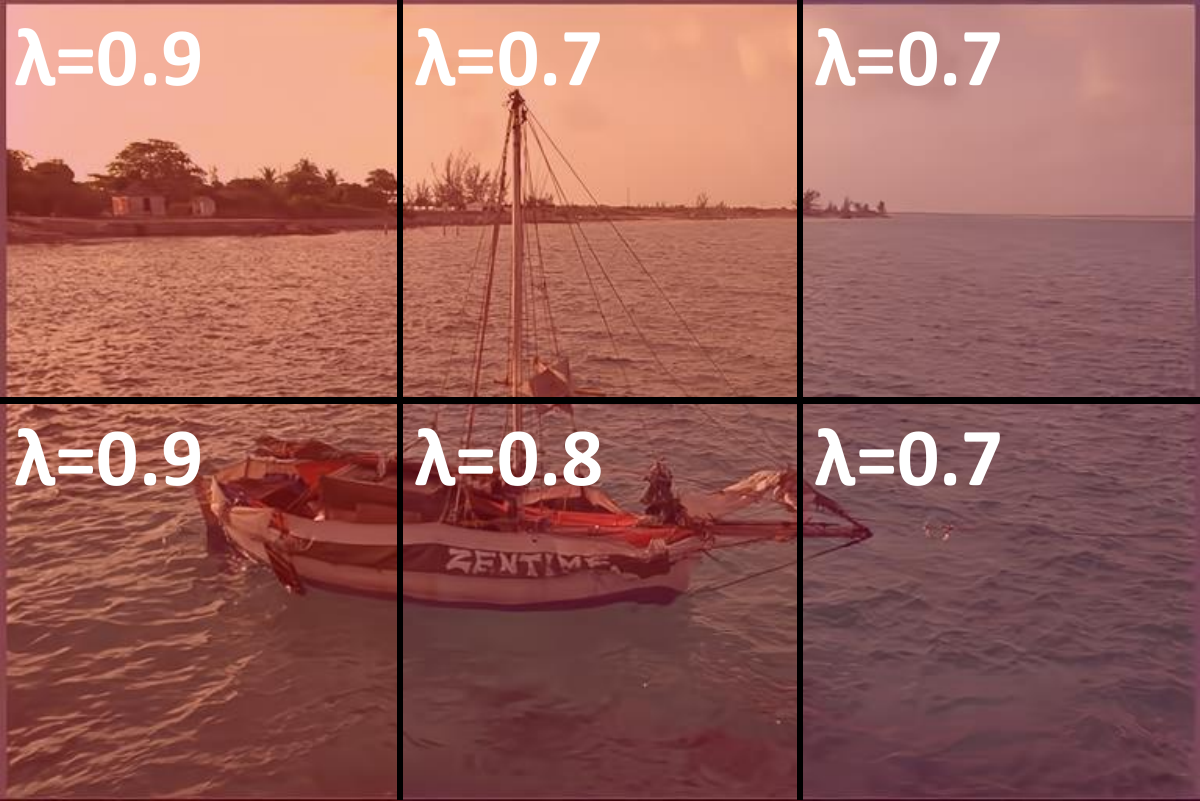}
            \end{minipage}
        }
        \subfigure[0.507bpp 32.40dB $\Delta R$=2.06\%]{
            \begin{minipage}[b]{0.4\textwidth}
            \label{fig:image_level_rate_bit_allocation}
            \includegraphics[width=1\textwidth]{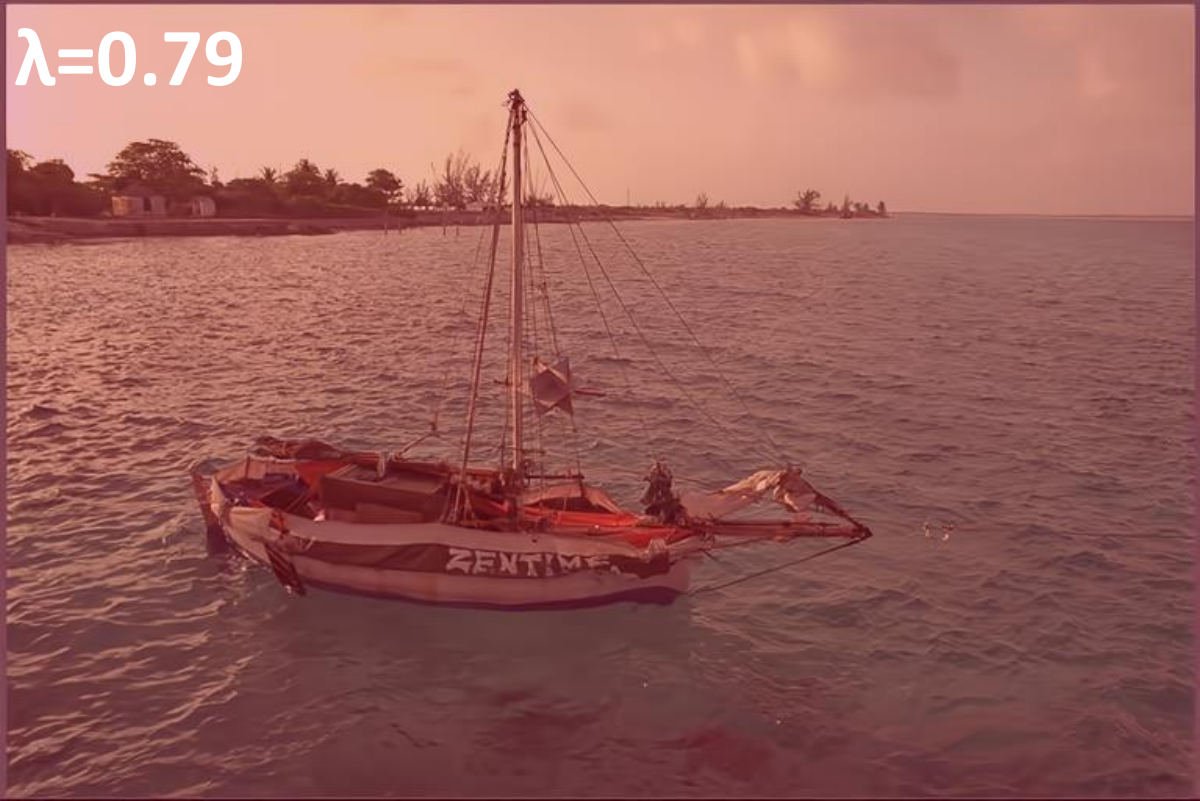}
            \end{minipage}
        }
        \subfigure{
            \begin{minipage}[b]{0.1\textwidth}
            \includegraphics[width=1\textwidth]{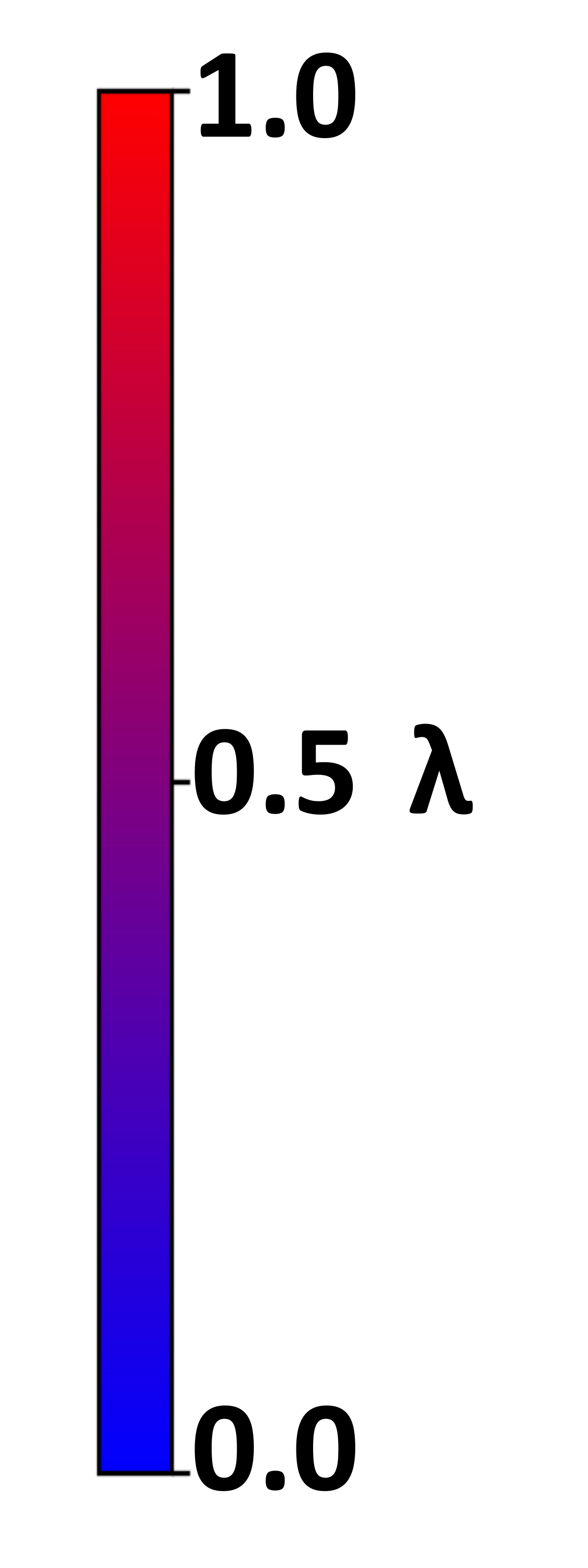}
            \end{minipage}
        }
	\caption{Comparison of block-level rate control results and image-level rate control results on kodim06. The block-level rate control is more accurate and the compression performance is comparable to that of the image level.}
 \label{fig:rate_control_bit_allocation}
\end{figure}

Figure~\ref{fig:block_level_rate_bit_allocation} and Figure~\ref{fig:image_level_rate_bit_allocation} illustrate the results of the bit allocation at respective block level and image level. It is evident that varying bitrates are assigned to image blocks with different contents when rate control is implemented at the block level. Image blocks with simple textures are allocated with lower bitrates, while those with complex textures are with higher bitrates. On the other hand, the rate control algorithm at the image level~\cite{jia2022rate} can only decide the single $\lambda$ value for the entire image without considering the texture complexity of different regions. In addition to improved compression performance for block-level coding by optimal bit allocation, coding in blocks also reduces the memory consumption, with only half of the image-based algorithms, which also offers potential for practical applications. 

\subsection{The Effect of Sampling Ratio}

To implement the rate control, we sample a certain percentage of blocks, compute their $R-D$ models, and predict the remaining unsampled blocks. In this process, the choice of sampling ratio has a significant impact on both the accuracy and speed of rate control. In order to evaluate the impact of different sampling ratios, we select different sampling ratios as an ablation study. The results are shown in Table~\ref{table:rate control with different ratio results}.

\begin{table}[htb]
\setlength{\tabcolsep}{3pt}
\caption{Rate control results with different sampling ratios.} 
\label{table:rate control with different ratio results}
\centering
\begin{tabular}{cccccccc}
\hline
 \multirow{2}{*}{Dataset}& \multirow{2}{*}{Ratio}& \multicolumn{2}{c}{low bpp} & \multicolumn{2}{c}{middle bpp} & \multicolumn{2}{c}{high bpp} \\
 &  & $\Delta$R(\%)  & Time(s)  & $\Delta$R(\%)      & Time(s) & $\Delta$R(\%)    & Time(s) \\ \hline
\multirow{3}{2cm}{\centering Kodak \\ (512$\times$768)}   & 1:1      & {0.29}  &{2.333}  & {0.74}  & {2.512} & {0.50}& {2.400}  \\
                         & 1:2      & 0.55  &1.146  & 0.77  & 1.214  & 0.36&1.192   \\
                         & 1:3      & 1.22  &{0.845}  & 1.64  & {0.844}  & 1.32&{0.832}   \\ \hline
\multirow{6}{2cm}{\centering Tecnick \\ (1200$\times$1200)} & 1:1      & {0.66}  &9.516  &{0.24} & 9.473 & {0.25}&10.017    \\
                         & 1:2      & 1.02  &4.953 & 0.68 & 5.044 & 0.69&5.459    \\
                         & 1:3      & 1.12  &3.602  & 0.67 & 3.636& 0.70& 3.878   \\
                         & 1:4      & 1.20  &2.820  &0.79  & 2.903  & 0.80&3.069    \\
                         & 1:6      & 1.27  &2.001  &0.80  & 2.071  & 0.74  & 2.226   \\
                         & 1:8      &{1.18}  &{1.627}  & {0.88}  & {1.786}  &{0.72}  & {1.793}  \\ \hline
\end{tabular}
\end{table}

The experimental results indicate that the more sampling blocks, the slower speed but the more accurate the rate control behaves. Conversely, reducing the number of sampling blocks results in a slight degradation in rate control performance, but significantly accelerate the running speed. On the Kodak dataset, the sampling ratios of 1:2 and 1:3 reduce accuracy by an average of 0.05\% and 0.88\% compared to the 1:1 sampling ratio, but increase speed by an average of 2.04 and 2.87 times. For the Tecnick dataset, transitioning from a 1:1 sampling ratio to a 1:8 sampling ratio reduces the average rate control accuracy from 0.383\% to 0.927\%. However, this results in an approximately 5.57$\times$ decrease in average processing speed.

\section{Conclusion}
\label{Conclusion}
In this paper, we propose an accelerated rate control algorithm that greatly improves the speed while maintaining high accuracy. We define a novel $D-\lambda$ function specifically for LIC taking into account the normalized Lagrange factor $\lambda$. Then we exploit model correlations between image blocks with the aid of average gradient to speed up the runtime. Experiments have shown that our best model can achieve more than 99\% accuracy and our fastest model can achieve up to 100$\times$ acceleration while maintaining over 98\% accuracy. However, the compression performance of block-level coding is still inferior than that of image-level coding even with the support of our more precise rate control. How to utilize inter-block correlation and use it in both coding and bit allocation is a direction we are committed to address in the future.

\section{References}
\bibliographystyle{IEEEbib}

\end{document}